\documentclass[a4paper,fleqn,usenatbib]{mnras}

\usepackage[T1]{fontenc}
\usepackage{ae,aecompl}

\usepackage{graphicx}	
\usepackage{amsmath}	
\usepackage{amssymb}	
\usepackage{color}
\usepackage{mathptmx} 
\usepackage{multirow}
\usepackage{wasysym}
\usepackage[table]{xcolor}
\usepackage{array}
\usepackage{caption}
\usepackage{multicol}
\usepackage{url}
\usepackage[figuresright]{rotating}
\usepackage{pdflscape}

\usepackage{txfonts}
\def\aj{AJ}  
\def\apj{ApJ}
\def\apjl{ApJ}
\def\apjs{ApJS}
\def\ao{Appl.~Opt.}
\def\aap{A\&A}
\def\icarus{Icarus}
\def\mnras{MNRAS}
\def\nat{Nature}
\def\nat{Nature}
\def\bain{Bull.~Astron.~Inst.~Netherlands}

\definecolor{gray}{gray}{0.62}
\definecolor{darkspringgreen}{rgb}{0.09, 0.45, 0.27}

\title[Dynamics of passing-stars-perturbed binary star systems]{Dynamics of passing-stars-perturbed binary star systems}

\author[Bancelin et al.]{
D. Bancelin$^{1}$\thanks{E-mail: david.bancelin@univie.ac.at}, T. Nordlander$^{2,3}$, E. Pilat-Lohinger$^{1}$
and B. Loibnegger$^{1}$
\\
$^{1}$Institute of Astrophysics, University of Vienna, T\"urkenschanzstrasse 17, A-1180 Vienna, Austria\\
$^{2}$Division of Astronomy and Space Physics, Department of Physics and Astronomy, Uppsala University, Box 516, 75120 Uppsala, Sweden\\
$^{3}$Research School of Astronomy and Astrophysics, The Australian National University, Canberra, ACT 2511, Australia\\
}

\date{Accepted XXX. Received YYY; in original form ZZZ}

\pubyear{2015}

\begin{document}
\label{firstpage}
\pagerange{\pageref{firstpage}--\pageref{lastpage}}
\maketitle

\begin{abstract}
In this work, we investigate the dynamical effects of a sequence of close encounters over 200 Myr varying in the interval of 10000 -- 100000 au between a binary star system and passing stars with masses ranging from 0.1$M_{\odot}$ to 10$M_{\odot}$. We focus on binaries consisting of two Sun-like stars with various orbital separations $a_{\scriptscriptstyle 0}$ from 50 au to 200 au initially on circular-planar orbits. We treat the problem statistically since each sequence is cloned 1000 times.\\
Our study shows that orbits of binaries initially at $a_{\scriptscriptstyle 0}$ = 50 au will slightly be perturbed by each close encounter and exhibit a small deviation in eccentricity (+0.03) and in periapsis distance (+1 and -2 au) around the mean value. However increasing $a_{\scriptscriptstyle 0}$ will drastically increase these variances: up to +0.45 in eccentricity and between +63 au and -106 au in periapsis, leading to a higher rate of disrupted binaries up to 50\% after the sequence of close encounters. Even though the secondary star can remain bound to the primary, $\sim$20\% of the final orbits will have inclinations greater than 10$^{\circ}$. As planetary formation already takes place when stars are still members of their birth cluster, we show that the variances in eccentricity and periapsis distance of Jupiter- and Saturn-like planets will inversely decrease with $a_{\scriptscriptstyle 0}$ after successive fly-bys. This leads to higher ejection rate at $a_{\scriptscriptstyle 0}$ = 50 au but to a higher extent for Saturn-likes (60\%) as those planets' apoapsis distances cross the critical stability distance for such binary separation.
\end{abstract}




\section{Introduction}\label{S:intro}

In our Galaxy, more than 1200 open clusters have been catalogued with masses typically between $10$ to $10^{\scriptscriptstyle 4}\,M_{\odot}$ and a lifetime between some million to billion years \citep{murdin01}. Amongst the most famous open clusters, we mention the Pleiades, the nearest star cluster from the Earth which is easily observable by naked eye. It is commonly accepted that stars form in clusters \citep{lada93} and in general they remain part of a cluster for at least 10$^{\scriptscriptstyle 8}$ years \citep{kroupa95,kroupa98}. Therefore, planetary systems or proto-planetesimal disks will suffer from gravitational perturbations due to stellar encounters in the early stage of a system. \\
Planetary formation already takes place when stars are still bound to their parent cluster. Indeed, \cite{brucalassi14,brucalassi16} identified several hot Jupiters orbiting stars in the dense open cluster $M67$ and \cite{meibom13} reported observations of transits of two G2V stars by planets smaller than  Neptune in the billion-year-old $NGC6811$ open cluster. They concluded that planetary formation is not prohibited in a dense cluster environment as small planets can survive. However, in the early stage of planetary formation, a planetesimal disk can be severely truncated as shown in \cite{kobayashi01}. These authors pointed out that an initial planetesimal disk extending up to 0.8 times the stellar encounter distance (typically 150 -- 200 au) and perturbed by passing stars, will lead to a planet-forming region limited to 40 -- 60 au, i.e. less than half the initial size of the disk. When giant planets are formed, \cite{spurzem09} and \cite{fuentes97} reported that fly-bys can increase the orbital eccentricity and inclination of planets, close encounters that can therefore shrink the size and number of members of planetary systems. According to \cite{fragner09}, close encounters smaller than 150 au, can also significantly increase the mass and semi-major axis of forming giant planets. \cite{laughlin98} pointed out that orbital disruptions can occur when Jovian planets interact with binary stars.\\
Our Sun is likely to have been formed in a cluster with an initial mass $\ge\,500\,M_{\odot}$ \citep{weidner04} but dissolved long time ago. However, some siblings of the Sun orbit around the Galactic center at a distance $\sim$ 100 pc from us and could be identified by accurate orbital and physical measurements \citep{portegies09}. The perturbation from close encounters on the dynamical and compositional structure of a protoplanetary disk is such that the planetesimal distribution can remain imprinted with this signature over most of the main-sequence lifetime of the star. Indeed, \cite{ida00} showed that stellar encounters of pericenter distances between 100 and 200 au could have pumped up the velocity distribution inside Neptune's 3:2 MMR allowing a more efficient capture of objects into the resonance during a phase of migration of the proto-Neptune. This could explain the high orbital eccentricities and inclinations of the Kuiper Belt objects. In a similar way, \cite{rickman04} highlighted also that fly-bys can explain the large periapsis distances observed in the scattered disk. Passing stars can also explain the injection of comets from the Oort cloud into observable orbits i.e. periapsis distances less than 5 au \citep{oort50, rickman76, rickman05,fouchard07}. As shown in \cite{rickman08} a huge number of comets from the Oort cloud can enter the observable region within 3 Gyr after the formation of the Sun. Moreover, \cite{fouchard11} emphasized the key role of massive stars which can increase the cometary flux by 40\%. Recently, \cite{nordlander17} investigated the survival of a primordial Oort Cloud, accounting for stellar cluster properties. Independently of the cluster mass, they concluded that an Oort Cloud can survive only when comets orbit at semi-major axes $\le$ 3000 au.\\
Many stars in our galaxy are members of binary systems and most of them must have formed as such \citep{goodman93}. Effects of an encounter between binary stars (initially at 10 au and on circular orbits) and a passing star have been studied by \cite{hills75}. They treated the problem statistically considering three family models depending on the mass ratio of the three stars. For each model, the pre-encounter velocity (from 0 -- 800 km/s) and impact parameter (0 -- 4 times the initial binary separation) were considered as constant. They showed that in addition to changes in the secondary's orbital elements, close encounters at zero impact parameter (i.e the encounter occurs at the binary's centre of mass) can either completely break the binary apart or enable the field star to be a stellar member (i.e. one of the original binary components has escaped to infinity).
\cite{heggie93} studied the effect of a fly-by on the secondary star's eccentricity in clusters assuming the encounter to be both tidal and slow. Their statistical results showed that the change of eccentricity declines in general as a function of a power law of the ratio of the binary's semi-major axis and the distance of the close encounter.\\

In our study, we aim to simulate the interactions between a binary star system and a stellar cluster. To this purpose, we investigate the dynamical evolution of binary stars gravitationally perturbed by a sequence of close encounters within 200 Myr. All simulations are performed with respect to the binary centre of mass as we neglect the gravitational perturbation of the cluster field as well as cluster tides. Our approach provides a statistical result as the passing star masses range from 0.1$M_{\odot}$ to 10$M_{\odot}$ and close encounter distances between 10000 -- 100000 au (with respect to the binary centre of mass). Sect. \ref{S:model} details the fly-by modelling and the numerical integration method of a binary star on the one hand and binary star systems hosting a gas giant planet on the other hand. Then, in Sect. \ref{S:results}, we present our statistical results of the perturbations on binary stars and then on a giant planet initially moving on a circular and planar orbit. Sect. \ref{S:conclusion} concludes our work.

\section{Methods}\label{S:model}

In this section, we first detail our method for the encounter modelling between a binary star system and interloping stars representing members of a loosely bound cluster. We select the initial mass of this cluster as $\mathcal{M}_{\scriptscriptstyle {cl}}$ = 1000\,$M_{\scriptscriptstyle \odot}$, which is a typical mass of known open clusters \citep{murdin01}. For each computation of a sequence of close encounters, the current mass of the cluster will be taken into account and only the most destructive close encounters will be used to assess the perturbation on the binary star.
        \begin{figure}
         \centering{
         \includegraphics[width=\columnwidth]{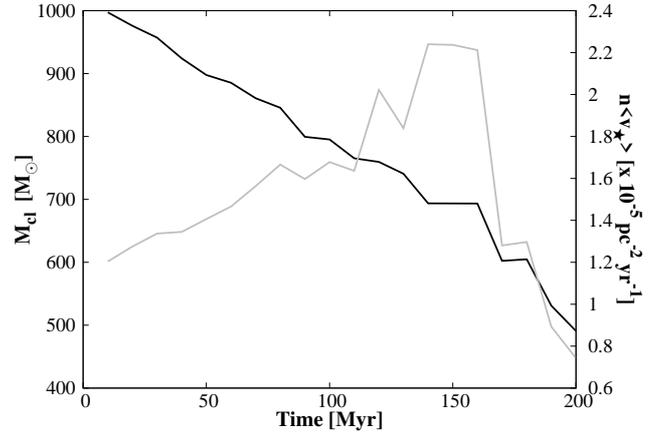}}
         \caption{Cluster mass $\mathcal{M}_{\scriptscriptstyle {cl}}$ (black) and encounter flux evolution $n<v_{\star}>$ (grey) with time computed at the half-mass radius.}\label{F:mass_cluster}
        \end{figure}
\subsection{Encounter modelling}\label{SS:enc}

To determine an encounter configuration, we proceed with the following 5-steps algorithm:
\begin{enumerate}
\item [(1)] We pre-compute a sample of encounters relative to the binary barycentre with the corresponding characteristics:\\

\begin{itemize}
\item the mass $m_{\star}$ of the passing star is chosen between $0.1\,M_{\odot}$ and $10M_{\odot}$ according to the initial stellar mass function as described in \cite{kroupa01} with a broken power law function:

\begin{equation}\label{E:mass1}
\Phi(m_{\star}) = \left\{
\begin{array}{l}
k_{\scriptscriptstyle 1}m_{\star}^{-\alpha_{1}} ~~~ m_0<m_{\star}\le m_1\\
k_{\scriptscriptstyle 2}m_{\star}^{-\alpha_{2}} ~~~ m_1<m_{\star}\le m_2
\end{array}
\right.
\end{equation}

where $\alpha_1 = 1.3$, $\alpha_2 = 2.3$, $m_0 = 0.1$M$_{\odot}$,  $m_1 = 0.5$M$_{\odot}$ and $m_2 = 150$M$_{\odot}$. $k_1$ and $k_2$ are normalization constants evaluated on the one hand at the limit mass of $\Phi(m_{\star})$ for $k_1$ and  by solving $X_1 + X_2 = 1$ for $k_2$ on the other hand, with:\\
$\displaystyle X_1 = \int_{m_0}^{m_1}\Phi(m_{\star})\,dm_{\star}$ ~~~~~ and ~~~~~  $\displaystyle X_2 = \int_{m_1}^{m_2}\Phi(m_{\star})\,dm_{\star}$

Finally, the mass $m_{\star}$ is deduced from a random number $\xi\, \in\, [0;1]$ such that:

\begin{equation}
\left\{
\begin{array}{l}
\displaystyle ~\mbox{if}  ~~0 \le \xi \le X_1 ~~\mbox{then}~~ \int_{m_0}^{m_{\star}}\Phi(m_{\star})dm_{\star} \Rightarrow m_{\star}\\
\displaystyle ~\mbox{if}  ~~ X_1 \le \xi \le X_1 + X_2  ~~\mbox{then}~~ X_1 +  \int_{m_1}^{m_{\star}}\Phi(m_{\star})dm_{\star} \Rightarrow m_{\star}
\end{array}
\right.
\end{equation}
   \begin{figure*}
     \centering{
     \begin{tabular}{cc}
          \includegraphics[width=0.5\textwidth]{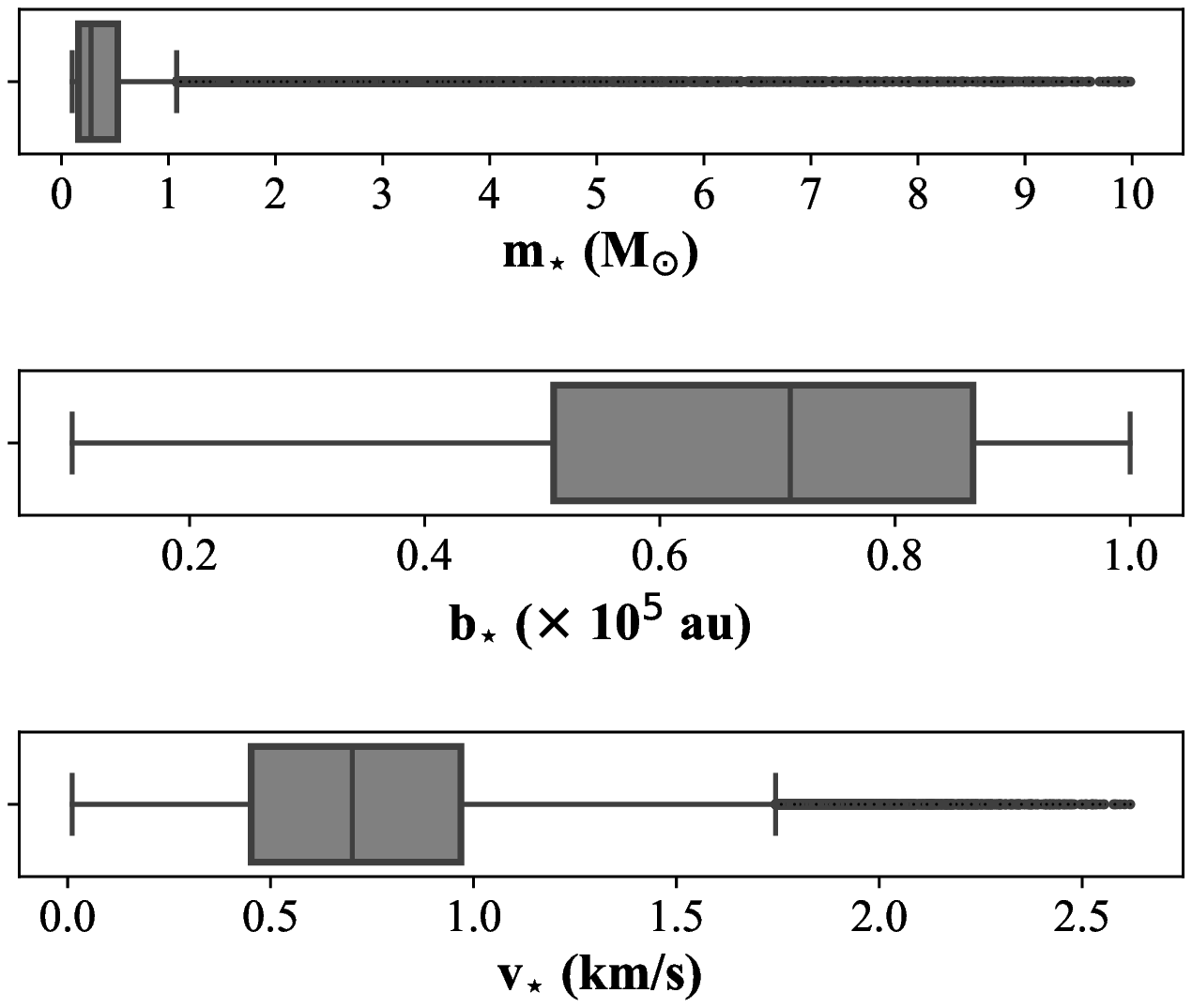} &
          \includegraphics[width=0.5\textwidth]{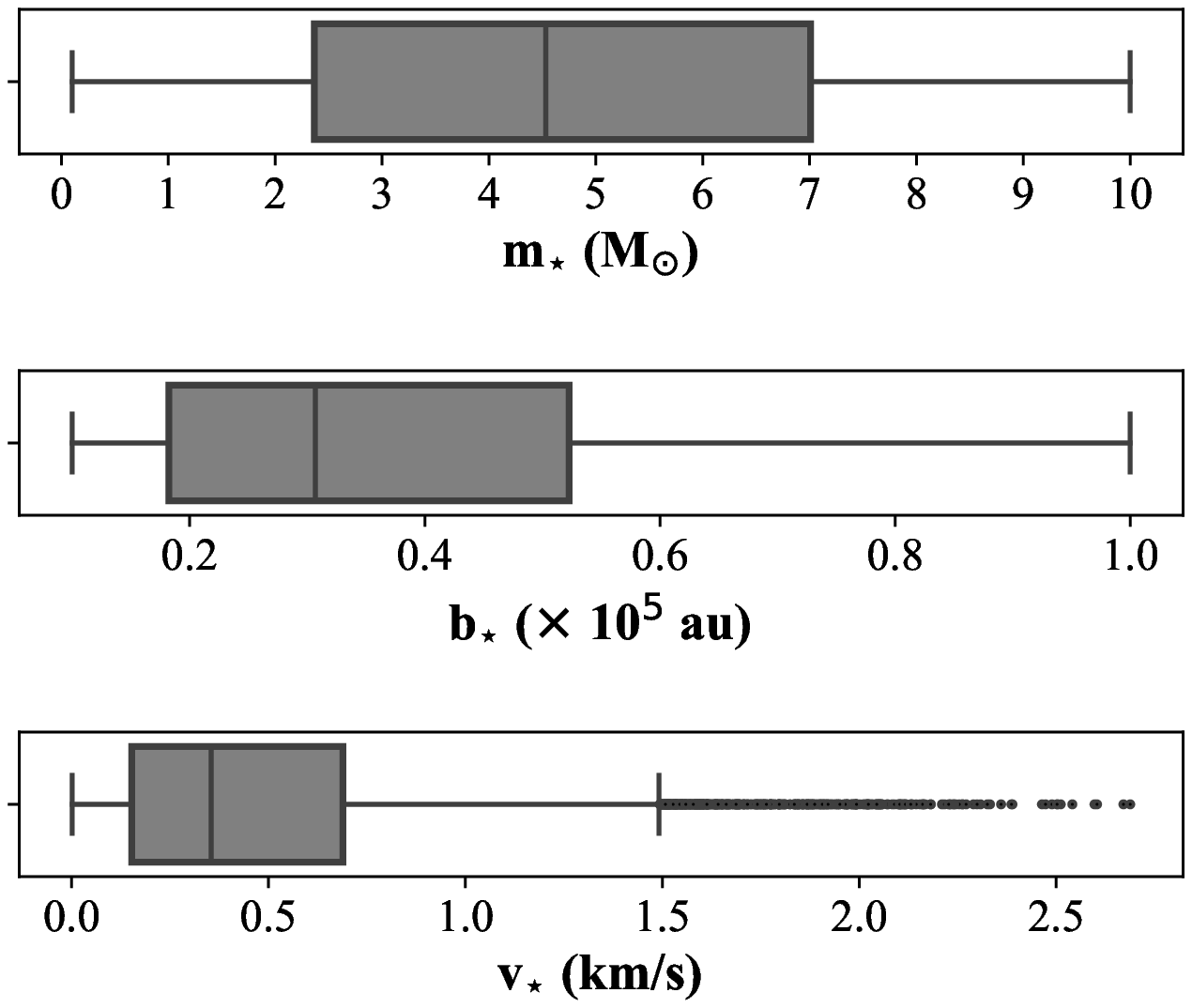}\\
     \end{tabular}}
     \caption{Boxplots showing the distributions in $m_{\star}$ (top), $b_{\star}$ (middle) and $v_{\star}$ (bottom) for the parent distribution of all expected encounters (left) computed from Eqs. (\ref{E:mass1}) -- (\ref{E:vel}), or the most important encounters that are actually executed (right) in Eq. \ref{Eq:param}. The straight vertical line inside the grey area is for the median value and the outliers are represented by dots outside the edges of the boxplot.} \label{F:distrib}
    \end{figure*}
As compact remnants such as white dwarfs, neutron stars and black holes likely form with a significant kick that should eject them from a loosely bound cluster, it is unlikely that they would be present in our sample during the timescale of our integration. We therefore identify these cases by evolving each star using the rapid stellar evolution code SSE \citep{hurley00} and only those evolving into one of the three remnant states mentioned above are removed and replaced by another one such that $0.1\,M_{\odot}\,\le\,m_{\star}\,<\,10\,M_{\odot}$.\\

\item we place the star in a cartesian coordinate system $\mathbf{d}_{\star}$ relative to the binary centre such as $|\mathbf{d}_{\star}|\, = b_{\star}$ where $b_{\star}$ is the impact parameter randomly chosen below $b_{\scriptscriptstyle {max}} = 100000$ au such that:
\begin{equation}
\displaystyle b_{\star} = \sqrt{\xi}\,b_{\scriptscriptstyle {max}} \mbox{ }\, \ge\, 10000 \mbox{ } au,  \mbox{ } \xi \in\, [0;1]
\end{equation}

\item we construct $\mathbf{d}_{\star}$ by randomly choosing its longitude $\alpha\, \in\, [0:2\pi]$ and its latitude $\delta$ such that $\sin\delta\, \in\, [-1;1]$.\\

\item to derive the relative velocity $v_{\star}$ with respect to the binary barycentre, we determine the tangential and radial velocity of the barycentre and passing star assuming a flat distribution of angular momentum \citep{king66}, in a similar way as described in \cite{nordlander17}.\\
At the closest approach, the relative velocity $\mathbf{v}_{\star}$ and position $\mathbf{d}_{\star}$ vectors are perpendicular. In order to find the velocity components, we have to solve the following system:
\begin{equation}\label{E:vel}
 \left\{
 \begin{array}{l}
\displaystyle  V_{\scriptscriptstyle \text{x}}^2 + V_{\scriptscriptstyle \text{y}}^2 + V_{\scriptscriptstyle \text{z}}^2 = 1  \\
\displaystyle  x\,V_{\scriptscriptstyle \text{x}} + y\,V_{\scriptscriptstyle \text{y}} + z\,V_{\scriptscriptstyle \text{z}} = 0
 \end{array}
 \right.
\end{equation}
with $V_{\scriptscriptstyle \text{i}} = v_{\scriptscriptstyle \text{i}}/ v_{\star}$ (i=x,y,z) and $x,y,z$ being the position coordinates. One can randomly choose one of the three velocity components between [-1;+1] and solve a second order equation to find the two remaining ones.\\
\end{itemize}

\item [(2)] The first step of the algorithm is iterated until $\sum\,m_{\star}\, \ge\, \mathcal{M}_{\scriptscriptstyle {cl}}(t)$ where $\mathcal{M}_{\scriptscriptstyle {cl}}(t)$ is the total mass of the cluster according to its age. The evolution and structure of the cluster was computed following \citet{nordlander17}. Briefly, we used the analytic code EMACSS \citep{alexander12,gieles14,alexander14} to compute the evolution of the cluster, and rescale its lifetime according to the predicted effects of GMC encounters \citep{gieles06}. We fit \citet{king66} models to the predicted mass, half-mass radius and core radius of the cluster every 10\,Myr of its evolution. As shown in Fig.~\ref{F:mass_cluster}, the cluster evolves rapidly and loses nearly half of its mass within the 200\,Myr period considered here. We also show the expected encounter rate at the half-mass radius, which itself varies with time.
 
\item [(3)] We introduce the parameter $S$ as the approximate impulse which would be transferred to the binary star
\begin{equation}\label{Eq:param}
S = \frac{m_{\star}}{v_{\star}\,b_{\star}}
\end{equation}
and we will select among all the pre-computed encounters, the one with the largest $S$ value to account for the single most important encounter.\\
In Fig. \ref{F:distrib} we show the statistics as boxplots of the pre-computed sample of encounters (left) for $m_{\star}$, $b_{\star}$ and $v_{\star}$ as determined from Eqs. (\ref{E:mass1}) -- (\ref{E:vel}). The extreme borders of these boxplots are for the minimum and maximum values of the data excluding outliers which are represented with dots symbols beyond the maximum value. Inside the grey areas is indicated the median value (vertical line) and their edges represent the 25$^{\scriptscriptstyle {th}}$ and 75$^{\scriptscriptstyle {th}}$ percentiles of the data respectively, meaning that 50\% of the pre-computed passing stars will have masses\footnote{The $m_{\star}$ distribution exhibits many outliers that are overlapping beyond 1$M_{\odot}$} between 0.1 -- 0.5$M_{\odot}$, impact parameters in the range of 0.5 -- 0.9$\times$10$^{\scriptscriptstyle 5}$ au and velocities within 0.5 -- 1.0 $km.s^{\scriptscriptstyle -1}$. Maximising the parameter $S$ from Eq. (\ref{Eq:param}) will change the statistics (Fig. \ref{F:distrib}, right) such that it will instead favour 50\% of the values of $m_{\star}$ to lie between 2.5 -- 7$M_{\odot}$, of $b_{\star}$ between 0.2 -- 0.5$\times$10$^{\scriptscriptstyle 5}$ au and of $v_{\star}$ between 0.1 -- 0.7$km\,s^{\scriptscriptstyle -1}$. \\ 

\item [(4)] In order to take into account the whole perturbation of the passing star during its motion, we assume that the star will move on a straight line with a constant velocity $v_{\star}$  \citep{rickman05,fouchard11} from an initial state $\mathbf{d}_{\star}^0$ such that $\left |\mathbf{d}_{\star}^{\scriptscriptstyle 0} \right|\, = b_{\scriptscriptstyle {max}}$. The position vector $\mathbf{d}_{\star}$ is thus linearly propagated backwards such that $\displaystyle \mathbf{d}_{\star}^0 = \mathbf{d}_{\star} - \mathbf{v}_{\star}t_{\scriptscriptstyle d}$ where $\displaystyle t_{\scriptscriptstyle d} = \sqrt{\frac{b_{\scriptscriptstyle {max}}^2 - b_{\star}^2}{v_{\star}^2}}$ is the time needed to reach that position.

\item [(5)]Finally, the initial barycentric positions of the primary and secondary components are also propagated to the time before the encounter by the same amount of time $t_{\scriptscriptstyle {d}}$.

\end{enumerate}

\subsection{Numerical integration of binary star systems}\label{SS:num_bin}

\begin{figure}
         \centering{
         \includegraphics[width=\columnwidth]{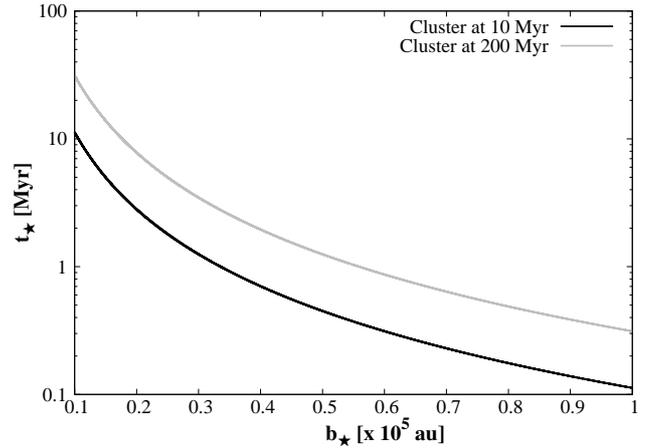}}
         \caption{The expected time, on average, between encounters within the impact parameter range $b_\star$ for the cluster at an age of 10\,Myr (black) and 200\,Myr (grey).}\label{F:tcol}
        \end{figure}

The first part of the simulations consists of integrating the binary star system which is sporadically perturbed by a star in the field. Because of the elliptic motion of a star with respect to the centre of its cluster, the encounter frequency will be highly dependent on its location \citep{nordlander17}. As we consider fly-bys with respect to the binary star barycentre we aren't able to account for this dependency. Instead, we follow the approach of \cite{nordlander17} by computing the time $t_{\star}$ separating two encounters according to the impact parameter $b_{\star}$. To do so, we took into account a typical value of the encounter flux $n<v_{\star}>$\footnote{$<v_{\star}>$ represents the average encounter velocities} at the half-mass radius varying with the age of the cluster as shown in Fig. \ref{F:mass_cluster} (grey curve). As one can observe, $n<v_{\star}>$ drops significantly around 150 Myr after reaching its maximum. Actually, the start condition is a quite "puffy" cluster, that contracts over about 100 Myr. Contraction reduces the half-mass radius, so the cluster becomes more compact and the encounter rate goes up. In other plots showing the evolution of the core radius with time, we observe that this parameter decreases very rapidly until it reaches a minimum, when the core has collapsed at ~150 Myr. After this, the encounter rate drops mainly because of mass loss, that reduces the average density in the cluster.\\
The time between two encounters is deduced from the inverse of the encounter rate, i.e. $\displaystyle t_{\star} = \frac{1}{nv_{\star}\,\pi\,b_{\star}}$. As seen in Fig. \ref{F:tcol} typical encounters occur more than once per 10\,Myr for a cluster of age 10\,Myr (black curve) and more than once per 30\,Myr for a cluster of age 200\,Myr (grey curve). As the selected distribution of $b_{\star}$ (Fig. \ref{F:distrib}, right) shows that $\sim$75\% of the interactions occur with impact parameters larger than 20\,000\,au, we therefore select a time between representative encounters $t_\star = 10$\,Myr. Fig.~\ref{F:tcol} clearly indicates that encounters at these distances are likely to occur at least this often. Therefore, one sequence of close encounters every 10 Myr over 200 Myr will contain 20 encounters. This sequence is pre-computed before the simulations by applying the 5-steps algorithm described in Sect. \ref{SS:enc} every 10 Myr.\\
        \begin{figure*}
         \centering{
         \includegraphics[width=0.5\columnwidth, angle=-90]{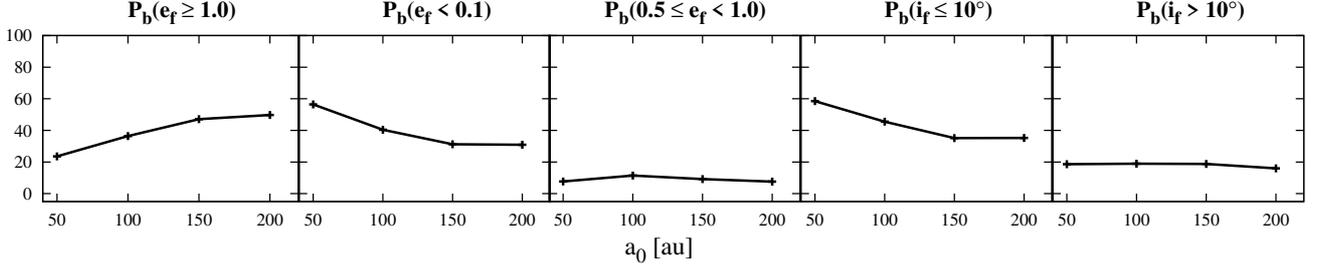}}
         \caption{Probability (in percent) of the secondary's final orbit according to its initial location $a_{\scriptscriptstyle 0}$.}\label{F:proba}
        \end{figure*}

  \begin{figure*}
     \centering{
     \begin{tabular}{cc}
          \includegraphics[width=0.5\textwidth]{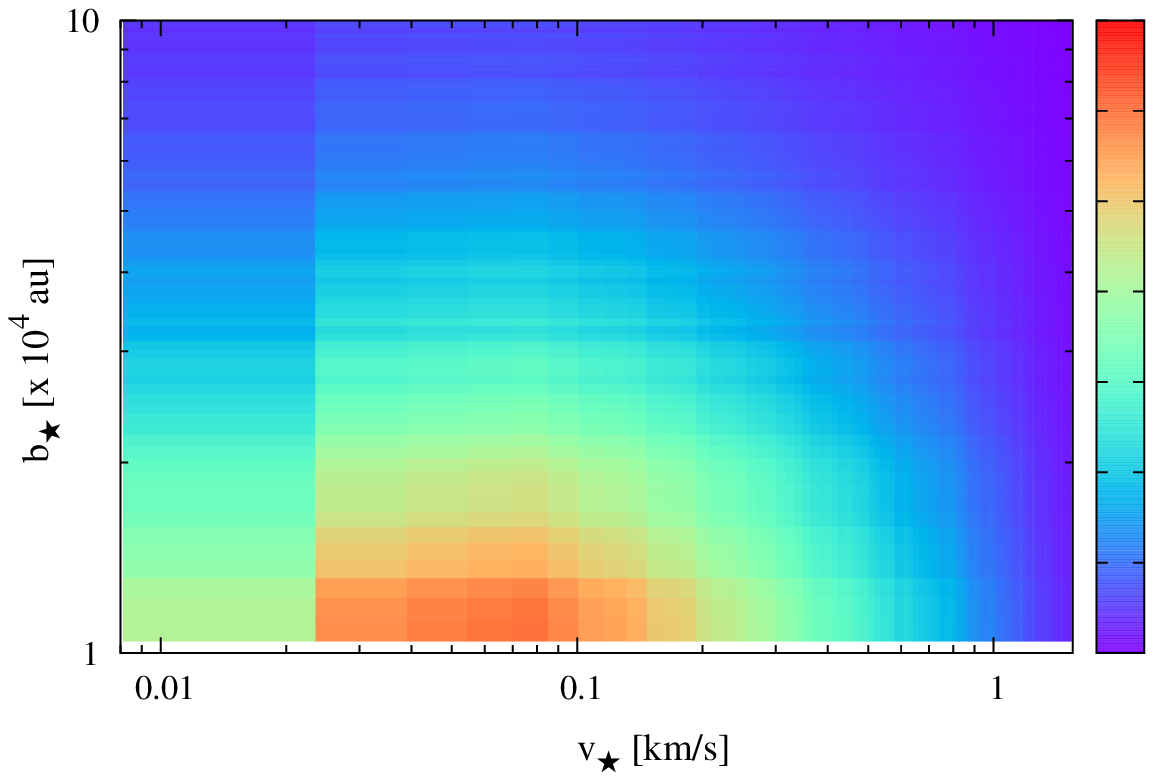} &
          \includegraphics[width=0.5\textwidth]{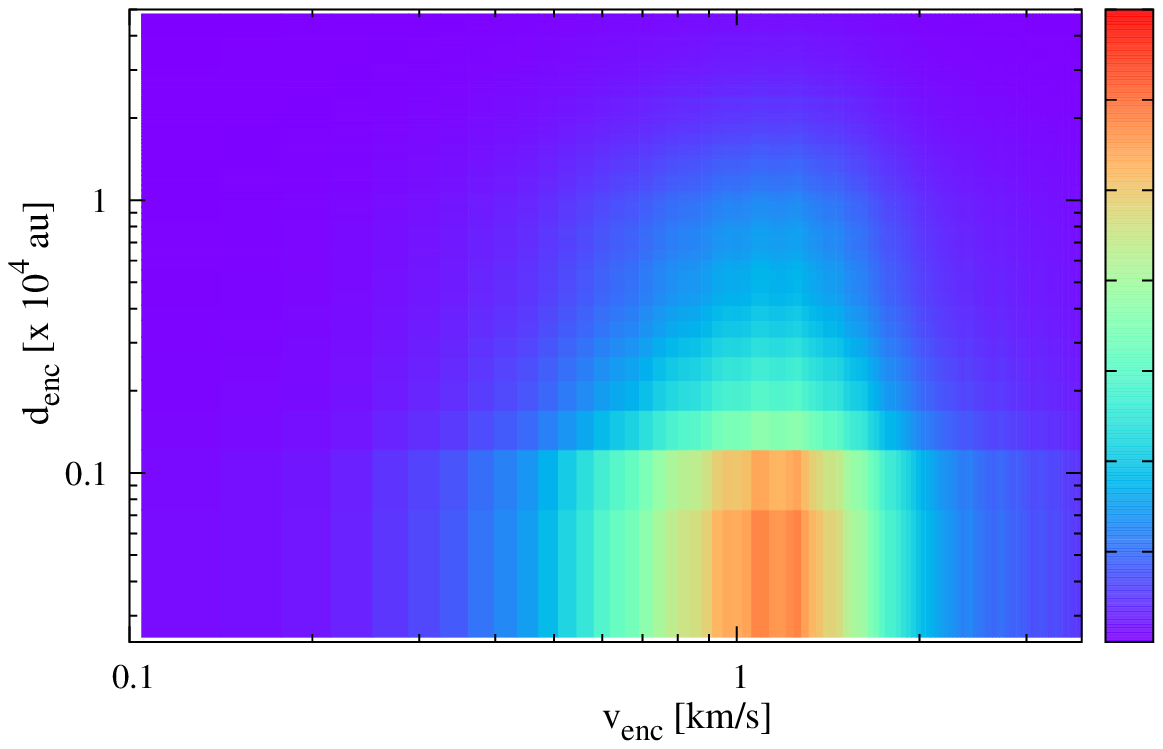}\\
     \end{tabular}}
     \caption{Left: distribution of the 2-body $v_{\star}$ and $b_{\star}$ as selected by the parameter $S$ in Eq. \ref{Eq:param}. Right: distribution of the encounter velocity and distance $v_{\scriptscriptstyle {enc}}$ and $b_{\scriptscriptstyle {enc}}$ between the passing star and binary's centre in a 3-body configuration.} \label{F:distrib_vit}
    \end{figure*}
We consider the circular-planar case for Sun-analogues binaries with initial separations of $a_{\scriptscriptstyle 0}$ = 50, 100, 150 and 200 au. Apart of the mean anomaly of the secondary star randomly chosen between 0$^{\circ}$ and 360$^{\circ}$, the initial orbital inclination \footnote{with respect to the initial orbital plane formed by the binary stars}, argument of periapsis and longitude of ascending node are all equal to 0$^{\circ}$.\\ 

We use a semi-analytical method to integrate the motion of a binary star system and the perturbation of passing stars, over 200 Myr in a barycentric frame:
\begin{enumerate}
	\item before and after a close encounter, we use the two-body problem to determine the positions of the stars,
	\item during close encounters we use the Lie integrator \citep{hanslmeier84,bancelin12} from the $nine$ package \citep{eggl10} to integrate numerically the motion of the three stars. As we want to study the stellar perturbation until the barycentric distance of the passing star theoretically reaches $b_{\scriptscriptstyle {max}}$ again, we integrate the three bodies over a timescale of $2\,t_{\scriptscriptstyle d}$\footnote{Because $t_{\scriptscriptstyle d}$ is defined by the linear motion approximation of the passing star, the latter will actually be located much beyond $b_{\scriptscriptstyle {max}}$ after the time $2\,t_{\scriptscriptstyle d}$ due the deviation during an encounter}.
\end{enumerate}

After each encounter, the current orbital parameters of the stellar companion, namely $a_{\scriptscriptstyle \text{b}}$, $e_{\scriptscriptstyle \text{b}}$, $i_{\scriptscriptstyle \text{b}}$ and $q_{\scriptscriptstyle \text{b}}$  are stored to derive statistical results. Here $q_{\scriptscriptstyle \text{b}}$ is the secondary's periapsis distance.\\

Finally, for each value $a_{0}$ studied we repeat 1000 times the procedure described in Sect. \ref{SS:enc}.

\subsection{Perturbations on gas giant planets}

In a second step, we investigate the perturbations induced by fly-bys on a gas giant planet initially on a circular-planar orbit at either 5.2 au with a Jupiter mass or at 9.5 au with a Saturn mass. Such initial locations are of interest for planetary formation, since many studies investigated their influences on the dynamical stability of terrestrial planets \citep{pilat08}, the water transport from a circumprimary disk of asteroids towards the habitable zone \citep{bancelin16} as well as the location of secular resonances affecting the circumprimary habitable zone of binary star systems \citep{pilat16,bazso17}.\\
The integrations are done separately and differently as we will consider the orbital state of the binary star system after each encounter obtained from the numerical integrations performed in Sect. \ref{SS:num_bin} as initial conditions of a three-body problem integrated with the Lie integrator.

\section{Results}\label{S:results}

\subsection{Statistical interpretation: I. The intermediate orbit}\label{SS:int}

The statistics on the intermediate orbit i.e. the orbital state before the end of the integration (200 Myr) are made using the so-called quartile parameters commonly used in descriptive statistics. We define the series $\mathbf{P}_{\scriptscriptstyle \text{x}}$ =  ($a_{\scriptscriptstyle \text{x}}$, $e_{\scriptscriptstyle \text{x}}$, $i_{\scriptscriptstyle \text{x}}$, $q_{\scriptscriptstyle \text{x}}$) constructed according to the orbital state of either the secondary star or a gas giant after each encounter with a passing star over the 1000 clones of sequences of 20 encounters within 200 Myr. We compute the median values $\mathbf{\overline{P}}_{\scriptscriptstyle \text{x}}$ =  ($\overline{a}_{\scriptscriptstyle \text{x}}$, $\overline{e}_{\scriptscriptstyle \text{x}}$, $\overline{i}_{\scriptscriptstyle \text{x}}$, $\overline{q}_{\scriptscriptstyle \text{x}}$) together with the minimum ${\mathbf{P}_{\scriptscriptstyle \text{x}}}_{|\scriptscriptstyle \text{min}}$ and maximum ${\mathbf{P}_{\scriptscriptstyle \text{x}}}_{|\scriptscriptstyle \text{max}}$ absolute deviation from the median value. We use an outliers criteria rejection procedure defined in \cite{tukey77} and \cite{frigge89} based on the interquartile range $I_{\scriptscriptstyle \text{QR}}$ parameter which is defined as follows: if $Q_{\scriptscriptstyle \text{25}}(\mathbf{P}_{\scriptscriptstyle \text{x}})$, $Q_{\scriptscriptstyle \text{50}}(\mathbf{P}_{\scriptscriptstyle \text{x}})$ and  $Q_{\scriptscriptstyle \text{75}}(\mathbf{P}_{\scriptscriptstyle \text{x}})$ are respectively the 25$^{\scriptscriptstyle \text{th}}$, the 50$^{\scriptscriptstyle \text{th}}$ and the 75$^{\scriptscriptstyle \text{th}}$ percentiles of the data, $Q_{\scriptscriptstyle \text{50}}$ being the median value $\mathbf{\overline{P}}_{\scriptscriptstyle \text{x}}$, then $I_{\scriptscriptstyle \text{QR}}(\mathbf{P}_{\scriptscriptstyle \text{x}}) = Q_{\scriptscriptstyle \text{75}}(\mathbf{P}_{\scriptscriptstyle \text{x}}) - Q_{\scriptscriptstyle \text{25}}(\mathbf{P}_{\scriptscriptstyle \text{x}})$. Therefore, a data is defined as an outlier if it lies beyond the interval $\left [ Q_{\scriptscriptstyle \text{25}}(\mathbf{P}_{\scriptscriptstyle \text{x}})-1.5\times I_{\scriptscriptstyle \text{QR}}(\mathbf{P}_{\scriptscriptstyle \text{x}}):Q_{\scriptscriptstyle \text{75}}(\mathbf{P}_{\scriptscriptstyle \text{x}})+1.5\times I_{\scriptscriptstyle \text{QR}}(\mathbf{P}_{\scriptscriptstyle \text{x}}) \right]$. After removing the outliers, we compute the minimum and maximum deviations defined as:
\begin{equation}
\begin{array}{l}
{\mathbf{P}_{\scriptscriptstyle \text{x}}}_{|\scriptscriptstyle \text{min}} = \mathbf{\overline{P}}_{\scriptscriptstyle x} - Q_{\scriptscriptstyle \text{25}}(\mathbf{P}_{\scriptscriptstyle \text{x}})+ 1.5\times I_{\scriptscriptstyle \text{QR}}(\mathbf{P}_{\scriptscriptstyle \text{x}}) \\
{\mathbf{P}_{\scriptscriptstyle \text{x}}}_{|\scriptscriptstyle \text{max}} = Q_{\scriptscriptstyle \text{75}}(\mathbf{P}_{\scriptscriptstyle \text{x}})+1.5\times I_{\scriptscriptstyle \text{QR}}(\mathbf{P}_{\scriptscriptstyle \text{x}}) - \mathbf{\overline{P}}_{\scriptscriptstyle x}
\end{array}
\end{equation} 
\subsection{Statistical interpretation: II. The final orbit}

The statistics on the final orbit are made according to the final orbital elements of the secondary star or a gas giant at the end of the sequence of encounters. Based on their final eccentricity $e_{\scriptscriptstyle f}$ and inclination $i_{\scriptscriptstyle f}$, we derive the following orbital probabilities defining their final orbit such as:

 \begin{itemize}
 \item [(a)] hyperbolic i.e. $e_{\scriptscriptstyle \text{f}} \ge 1.0$;\\
 \item [(b)] nearly circular i.e. $e_{\scriptscriptstyle \text{f}} < 0.1$;\\
 \item [(c)] highly eccentric i.e. $0.5 \le e_{\scriptscriptstyle \text{f}} < 1.0$; \\
 \item [(d)] bound and planar i.e. $e_{\scriptscriptstyle \text{f}} < 1.0$ and $i_{\scriptscriptstyle \text{f}} \le 10^{\circ}$;\\
  \item [(e)]bound and inclined i.e. $e_{\scriptscriptstyle \text{f}} < 1.0$ and $i_{\scriptscriptstyle \text{f}} > 10^{\circ}$
 \end{itemize}
 
\subsection{Perturbations on a binary star system}

We compile in Tab. \ref{T:circular} the statistical results for the perturbed secondary star -- according to its initial separation $a_{\scriptscriptstyle 0}$ -- based on the descriptive statistics in Sect. \ref{SS:int}. The series $\mathbf{P}_{\scriptscriptstyle \text{b}}$ =  ($a_{\scriptscriptstyle \text{b}}$, $e_{\scriptscriptstyle \text{b}}$, $i_{\scriptscriptstyle \text{b}}$, $q_{\scriptscriptstyle \text{b}}$) is constructed by accounting for the orbital perturbation induced by each stellar encounter before the end o the integration. For each mean value $\overline{a}_{\scriptscriptstyle \text{b}}$, $\overline{e}_{\scriptscriptstyle \text{b}}$, $\overline{i}_{\scriptscriptstyle \text{b}}$ and $\overline{q}_{\scriptscriptstyle \text{b}}$ we indicate their upper and lower deviations.\\
One notices that despite the large encounter distances (see Fig. \ref{F:distrib}), a sequence of fly-bys can significantly modify the initial orbit of the secondary. Although the mean values ($\overline{a}_{\scriptscriptstyle \text{b}}$, $\overline{e}_{\scriptscriptstyle \text{b}}$, $\overline{i}_{\scriptscriptstyle \text{b}}, \overline{q}_{\scriptscriptstyle \text{b}}$) $\sim$ ($a_{\scriptscriptstyle 0}$, $e_{\scriptscriptstyle 0}$, $i_{\scriptscriptstyle 0}$, $q_{\scriptscriptstyle 0}$), the upper and lower deviations are significant enough to provoke high perturbations on planetesimal disks. Indeed, ${e_{\scriptscriptstyle \text{b}}}_{|\scriptscriptstyle \text{max}}$ can reach +0.04 for $a_{\scriptscriptstyle 0}$ = 50 au and +0.45 for $a_{\scriptscriptstyle 0}$ = 200 au leading to an initial $q_{\scriptscriptstyle 0}$ reduced by $\sim$ 4\% for $a_{\scriptscriptstyle 0}$  = 50 au and by $\sim$ 54\% or $a_{\scriptscriptstyle 0}$  = 200 au. Last but not least, variances are large in $i_{\scriptscriptstyle b}$ from +12$^\circ$ for $a_{\scriptscriptstyle 0}$  = 50 au to 24$^\circ$ for $a_{\scriptscriptstyle 0}$  = 200 au. This is a behaviour predicted by the Gauss perturbation equations as the perturbed inclination of the secondary is $\displaystyle \frac{di_{\scriptscriptstyle b}}{dt}\,\propto \frac{1}{\sqrt{1-e_{\scriptscriptstyle b}^2}}$ and any increase of $e_{\scriptscriptstyle b}$ due to the passing star will necessarily lead to an increase of $i_{\scriptscriptstyle b}$.\\
\begin{table}
 \begin{center}
  \caption{Statistics on the secondary's intermediate orbital elements ($\overline{a}_{\scriptscriptstyle \text{b}}$, $\overline{e}_{\scriptscriptstyle 
\text{b}}$, $\overline{i}_{\scriptscriptstyle \text{b}}$ and $\overline{q}_{\scriptscriptstyle \text{b}}$) according to its initial location $a_{\scriptscriptstyle 0}$. The median value of the distributions are given together with their minimum and maximum deviation (lower and upper script, respectively).}
  \label{T:circular}
  \begin{tabular}{*{5}{l}}
   \hline

 $a_{\scriptscriptstyle 0}$ [au]    & 50 & 100 & 150 & 200  \\
      
\hline

\multicolumn{1}{l}{$\overline{a}_{\scriptscriptstyle {\text{b}}}$ [au]}&  \multicolumn{1}{l|}{50\,$^{\scriptscriptstyle {+  0}}_{\scriptscriptstyle {-  0}}$}& \multicolumn{1}{l|}{100\,$^{\scriptscriptstyle {+ 2}}_{\scriptscriptstyle {- 4}}$}& \multicolumn{1}{l|}{150\,$^{\scriptscriptstyle {+ 5}}_{\scriptscriptstyle {- 8}}$}&
200\,$^{\scriptscriptstyle {+7}}_{\scriptscriptstyle {-11}}$\\
\multicolumn{1}{l}{$\overline{e}_{\scriptscriptstyle {\text{b}}}$}        & \multicolumn{1}{l|}{0.0\,$^{\scriptscriptstyle {+0.04}}_{\scriptscriptstyle {-0.00}}$}& \multicolumn{1}{l|}{0.00\,$^{\scriptscriptstyle {+0.36}}_{\scriptscriptstyle {-0.00}}$}& \multicolumn{1}{l|}{0.00\,$^{\scriptscriptstyle {+0.45}}_{\scriptscriptstyle {-0.00}}$}& 0.00\,$^{\scriptscriptstyle {+0.45}}_{\scriptscriptstyle {-0.00}}$\\
\multicolumn{1}{l}{$\overline{i}_{\scriptscriptstyle {\text{b}}}$ [$^\circ$]}  & \multicolumn{1}{l|}{0\,$^{\scriptscriptstyle {+ 13}}_{\scriptscriptstyle {-  0}}$}&  \multicolumn{1}{l|}{1\,$^{\scriptscriptstyle {+ 22}}_{\scriptscriptstyle {-  1}}$}&  \multicolumn{1}{l|}{2\,$^{\scriptscriptstyle {+ 27}}_{\scriptscriptstyle {-  2}}$}&
2\,$^{\scriptscriptstyle {+ 24}}_{\scriptscriptstyle {-  2}}$\\
\multicolumn{1}{l}{$\overline{q}_{\scriptscriptstyle {\text{b}}}$ [au]}        &  \multicolumn{1}{l|}{50\,$^{\scriptscriptstyle {+  1}}_{\scriptscriptstyle {-  2}}$}&  \multicolumn{1}{l|}{100\,$^{\scriptscriptstyle {+ 26}}_{\scriptscriptstyle {- 44}}$}& \multicolumn{1}{l|}{150\,$^{\scriptscriptstyle {+ 51}}_{\scriptscriptstyle {-85}}$}&
200\,$^{\scriptscriptstyle {+63}}_{\scriptscriptstyle {-106}}$\\
\hline
  \end{tabular} 
 \end{center}
 \end{table}
Figure \ref{F:proba} shows the statistics on the final orbital state of the binary. The middle panel $\mathcal{P}_{\scriptscriptstyle b}(0.5\,\le\,e_{\scriptscriptstyle \text{f}}< 1.0)$ reveals that it is very unlikely that the secondary ends in a highly eccentric orbit and the two possible outcomes are either the binary is dissociated (first panel) or the secondary remains in a nearly circular orbit (second panel)\footnote{A small fraction of the orbit will have $0.1\le\,e_{\scriptscriptstyle}\,<0.5$}. The likelihood of a secondary to remain bound to the primary drops by 20\% for $a_{\scriptscriptstyle 0}$ = 50 au and by 50\% for $a_{\scriptscriptstyle 0}$ = 200 au.\\
Such behaviour is due to the dynamics of the encounter itself. In Fig. \ref{F:distrib_vit} we represent a 2-body map of $v_{\star}$ and $b_{\star}$ (left) as determined by Eq. \ref{Eq:param} where all selected encounters should occur at low relative velocity. This ($v_{\star}$, $b_{\star}$) distribution corresponds to the linear approximation trajectory of the passing star with respect to the binary's barycentre. However, because of the mutual perturbation between the binary and the passing star in a 3-body configuration, the encounter distance $d_{\scriptscriptstyle {enc}}$ with the binary centre can be deeper. Therefore, the passing star would fly-by at a higher encounter velocity $v_{\scriptscriptstyle {enc}}$, as represented on the right side of Fig. \ref{F:distrib_vit}. As we can see, severe encounters down to $d_{\scriptscriptstyle {enc}} \le\,1000$ au can occur at higher expected velocities in the range $v_{\scriptscriptstyle {enc}}\,\in\,[0.5:2.0]$ km.s$^{\scriptscriptstyle {-1}}$. Finally, as predicted by Tab. \ref{T:circular}, the very few orbits reaching inclinations beyond $10^{\circ}$ will result in lower probabilities for binaries with final inclined orbits (last panel).

\subsection{Perturbations on gas giant planets}

\begin{figure*}
	\centering{
		\begin{tabular}{c}
			\includegraphics[width=\textwidth]{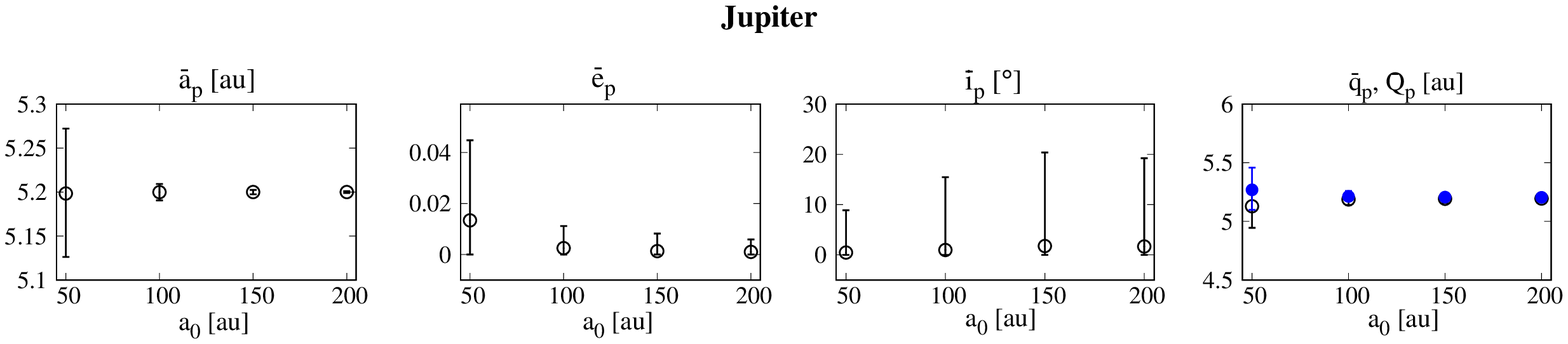}\\
			\includegraphics[width=\textwidth]{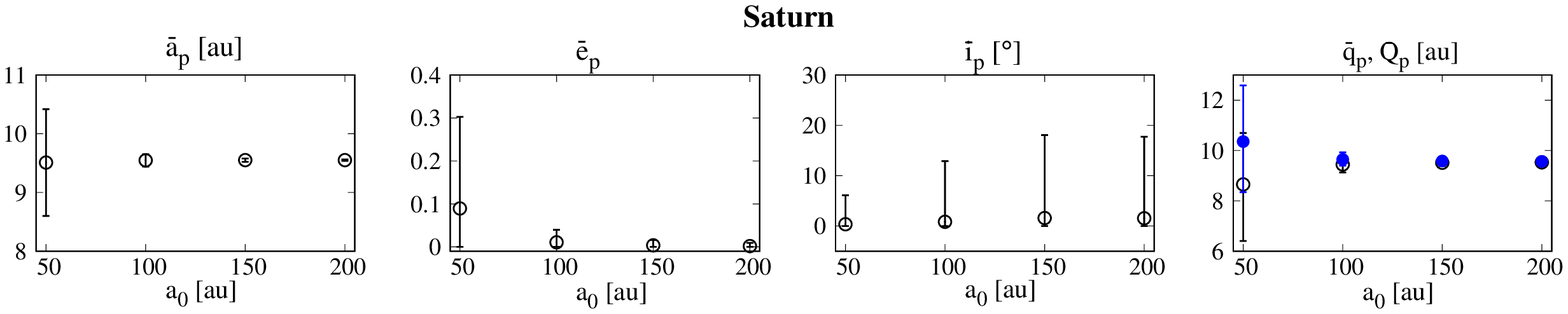}
	\end{tabular}}
	\caption{Orbital variations of a Jupiter (top) and a Saturn (bottom) gas giant. The patterns and error bars in blue are for the apoapsis distance of the planet.}\label{F:orbit_pla}
\end{figure*}

\begin{figure}
	\centering{
		\begin{tabular}{c}
			\includegraphics[width=\columnwidth]{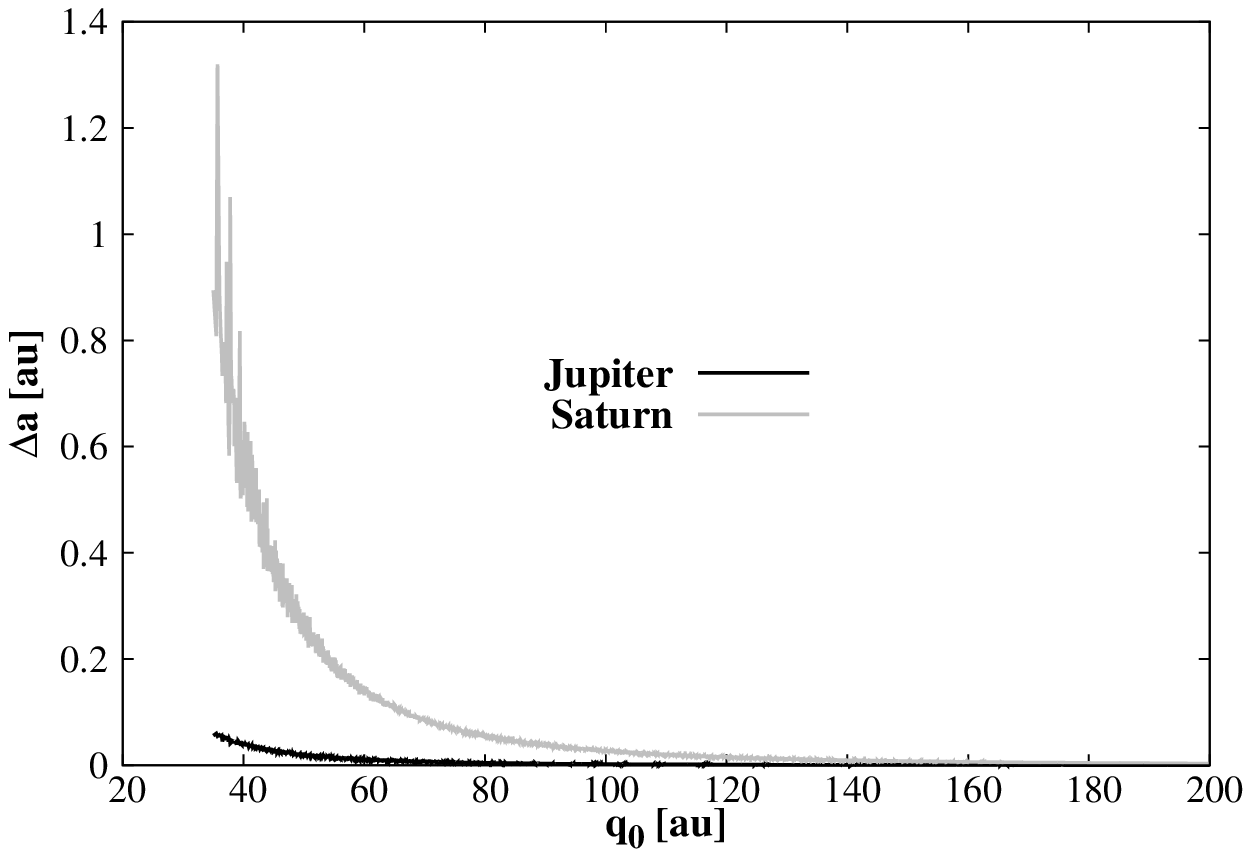} \\
			\includegraphics[width=\columnwidth]{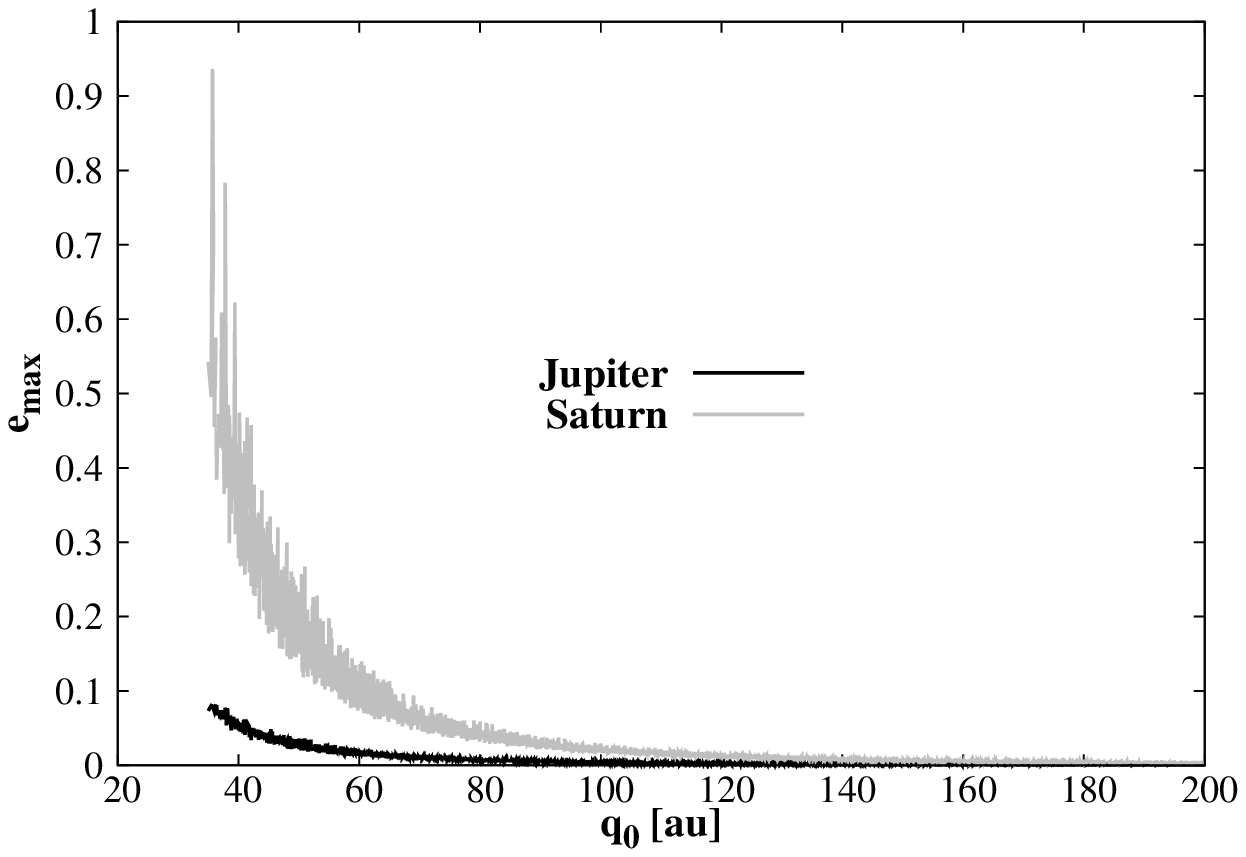} 
	\end{tabular}}
	\caption{Variations in semi-major axis (top) and maximum eccentricity of a Jupiter (black) and a Saturn (grey) with respect to the initial periapis $q_{\scriptscriptstyle 0}$ of the secondary star.}\label{F:orbit}
\end{figure}

\begin{figure*}
         \centering{
         \includegraphics[width=\textwidth]{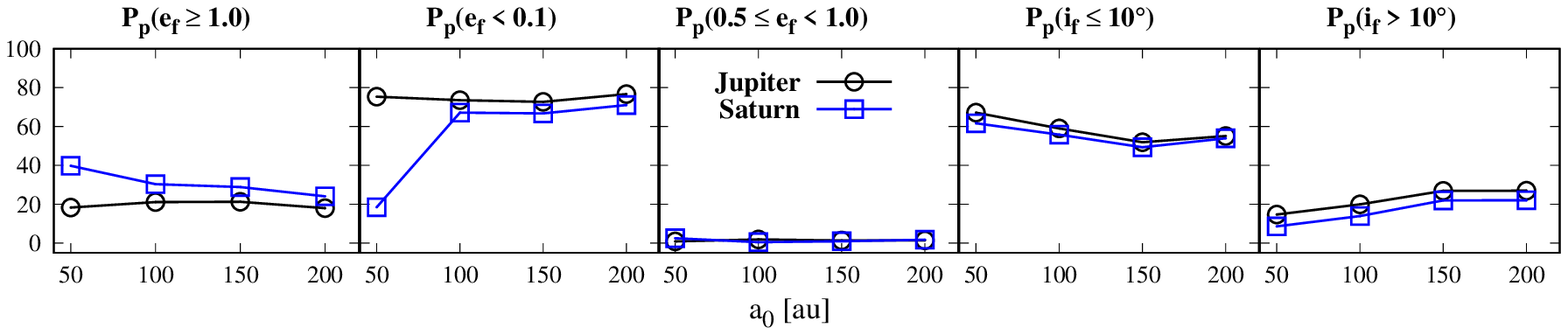}}
         \caption{Probability (in percent) of a Jupiter's (black) and a Saturn's (blue) final orbits according to the binary's initial location $a_{\scriptscriptstyle 0}$.}\label{F:proba_pla}
        \end{figure*}

Figure \ref{F:orbit_pla} shows results for the set of orbital elements $\mathbf{P}_{\scriptscriptstyle \text{p}}$ =  ($a_{\scriptscriptstyle \text{p}}$, $e_{\scriptscriptstyle \text{p}}$, $i_{\scriptscriptstyle \text{p}}$, $q_{\scriptscriptstyle \text{p}}$) for a Jupiter-like (top) and a Saturn-like (bottom) planet where we present the median value $\mathbf{\overline{P}}_{\scriptscriptstyle \text{p}}$ (open circle) together with the minimum and maximum values (error bars) of the $\mathbf{P}_{\scriptscriptstyle \text{p}}$ series. One important feature exhibited is the variations of both the semi-major axis (first panels) and the eccentricity (second panels) of the planets for tight binary separation. Indeed both gas giants' initial semi-major axes can drift upwards and downwards by an amount of $\sim$ 5.2\% for Jupiter and $\sim$ 10\% for Saturn. This is due to a dynamical characteristic already mentioned in \cite{bancelin16} for binary stars hosting a gas giant planet which can have its initial position periodically shifted by an amount $\Delta\,a$  which amplitude depends on the initial periapsis distance $q_{\scriptscriptstyle 0}$ of the secondary star. To highlight this feature, we performed independent simulations in which we varied the initial separation and eccentricity of the binary companion and investigated the perturbations in semi-major axis and eccentricity of a gas giant planet initially on a circular-planar orbit. As seen in the top panel of Fig. \ref{F:orbit}, the amplitude $\Delta\,a$ of a massive planet's semi-major is stronger for small initial $q_{\scriptscriptstyle 0}$ (top panel).\\
Another result from these independent simulations shown in Fig. \ref{F:orbit} (bottom panel) indicates an increase in eccentricity of the gas giant (also predicted by \cite{georgakarakos03}) for decreasing value of $q_{\scriptscriptstyle 0}$ (second panel of the top and bottom figures in Fig. \ref{F:orbit_pla}).\\
Especially for a binary's initial $a_{\scriptscriptstyle 0}$ of 50 au, small changes caused by a fly-by will translate into a change of the parameters for a gas giant orbiting the primary star. As a consequence, a Jupiter's periapis distance $q_{\scriptscriptstyle p}$ shows only small variations (Fig. \ref{F:orbit_pla}, upper figure, last panel) which moderately perturb embryos, planetary systems or asteroid rings orbiting in lower orbits, whereas a Saturn could nearly destroy any object in that region as its periapis can reach values down to 6 au.\\
Finally, the change in orbital inclination of the secondary also translates into a change of the initial orbital inclination of the gas giant and it can be forced to also reach similar maxima as for the binary.\\
For the final orbital state of a gas giant after a sequence of passing stars, Fig. \ref{F:proba_pla} depicts the orbital outcomes of a Jupiter (black) and a Saturn (blue). Despite the low rate of disrupted binaries at $a_{\scriptscriptstyle 0}$ = 50 au as shown in Fig. \ref{F:proba} one can notice that for that binary separation, the rate of ejected Saturns $\mathcal{P}_{\scriptscriptstyle b}(e_{\scriptscriptstyle \text{f}}\,\ge 1.0)$ $\sim$ 40\% is much higher than the rate of ejected Jupiters $\sim$ 20\% . This accounts for the critical semi-major axis \citep{pilat-lohinger02} which added up to its variance \citep{holman99} brings its minimum value down to $\sim$ 12 au for $a_{\scriptscriptstyle 0}$ = 50 au. Therefore, Saturn-like planets having apoapsis distances $Q_{\scriptscriptstyle p}\,\ge\, 12$ au (Fig. \ref{F:orbit_pla}) will inevitably be ejected. In spite of the high numbers of ejected binary for $a_{\scriptscriptstyle 0}\, \ge$ 100 au (Fig. \ref{F:proba}), giant planets (and therefore a planetary system in the HZ) could survive the loss of a binary companion as we see in Fig. \ref{F:proba_pla} that the rate of ejected gas giants for $a_{\scriptscriptstyle 0}\, \ge$ 100 au doesn't increase with the rate of disrupted binaries.\\
If we compare the probabilities $\mathcal{P}_{\scriptscriptstyle b}(0.5\,\le\,e_{\scriptscriptstyle \text{f}}< 1.0)$ and $\mathcal{P}_{\scriptscriptstyle b}(e_{\scriptscriptstyle \text{f}}< 0.1)$, one notices that Jupiter- and Saturn-like planets are likely to remain on nearly circular orbits for any value of $a_{\scriptscriptstyle 0}$ except for $a_{\scriptscriptstyle 0}$ = 50 au, location of which Saturns will suffer from strong perturbations as mentioned above.\\
As for the final inclinations (Fig. \ref{F:proba_pla} two last panels), most Jupiters and Saturns still orbiting around the primary despite the fly-bys will end up in orbits with inclinations less than 10$^\circ$ as their plane of orbital motion will be forced to follow the secondary's change of orbital inclination.  

\section{Conclusions}\label{S:conclusion}

In this paper, we investigated the influence of a sequence of passing stars on the motion of a binary consisting of two G2V stars with various orbital separations and initially moving on circular planar orbits. The sequence of passing stars has been modeled by randomly choosing the mass $m_{\star}$, the impact parameter $b_{\star}$ and velocity $v_{\star}$. We treated the problem statistically where we cloned 1000 times a sequence of fly-bys over 200 Myr. A key parameter of the dynamical orbital outcome is the initial separation $a_{\scriptscriptstyle 0}$ of the binary as we showed that placing the secondary further away from the primary will ineluctably lead to a high disruption rate of the system. As gas giant planets can already form while stars are still bound to their birth cluster, we showed that passing stars can indirectly affect the gas giants (e.g. Jupiter- or Saturn-like planets) quite strong through the binary's perturbed orbit. In our study, we highlighted two possible outcomes according to its initial orbital separation $a_{\scriptscriptstyle 0}$: 
\begin{itemize}
\item Either the semi-major axis and eccentricity of the giant planets oscillate after each fly-by -- strongly for a Saturn and moderately for a Jupiter -- for $a_{\scriptscriptstyle 0}$ = 50 au; as a consequence, the rate of ejections of Saturn-like planets is very high ($\sim\,40\%$) compared to the Jupiters ($\sim\,20\%$);
\item or the planet's semi-major axis and eccentricity remain almost unchanged for $a_{\scriptscriptstyle 0}\,\ge$ 100 au: as a consequence most gas giants will remain on nearly circular orbits. 
\end{itemize}

Important parameters are certainly the periapsis distance and inclination of the gas giant planet. Protoplanetary systems in the region of the HZ will undoubtedly be indirectly affected by the orbital change of the gas giant caused by fly-bys as shown in \cite{bitsch13}. For larger initial values of $a_{\scriptscriptstyle 0}$, a high secondary's inclination might not be critical for planetary formation as shown in \cite{batygin11}.\\
An important factor for the habitability of planets formed in the habitable zone is their water content which is strongly correlated to the gas giant and stellar companion's orbital and physical parameters \citep{haghighipour07,haghighipour09} leading to nearly dry planets in the HZ of the host star. However, a circumprimary icy asteroid belt can be the source of water for the entire habitable zone \citep{bancelin15,bancelin16} and for embryos or planets up to Earth distances \citep{bancelin17}. In a future work, we aim to study how direct or indirect perturbation from passing stars can influence the water transport to a circumprimary habitable zone from circumstellar or circumbinary disks. 

\section*{Acknowledgements}

DB, EPL and BL acknowledge the support of the Austrian Science Foundation (FWF) NFN project: Pathways to Habitability and related sub-projects S11608-N16 "Binary Star Systems and Habitability" and S11603-N16 "Water Transport". DB and EPL acknowledge also the Vienna Scientific Cluster (VSC project 70320) 
for computational resources.






\appendix

\bibliographystyle{mnras}
\bibliography{biblio} 
\bsp	
\label{lastpage}
\end{document}